 \documentclass[11pt]{article} \usepackage{geometry} 
   \usepackage[parfill]{parskip}
    \usepackage{graphicx}
     \usepackage{amssymb}
      \usepackage{amsmath}
       \usepackage{epstopdf}
	\usepackage{hyperref}
\usepackage{datetime}

\title{Dispersion measure: Confusion, Constants \&\ Clarity}
 \author{S.\ R.\ Kulkarni} 

\begin{document}
 \maketitle

\noindent{\bf \large Abstract.} The dispersion measure (DM) is one
of the key attributes of radio pulsars and Fast Radio Bursts (FRBs).
There is a mistaken view that the DM is an accurate measure of  the
column density of electrons between the observer and the source.
To start with, the DM, unlike a true column density, is not  a
Lorentz invariant. Next, the DM also includes contribution from
ions and is sensitive to the temperature of the plasma in the
intervening clouds.  Separately, the primary observable is  the
dispersion slope, $\mathcal{D}\equiv \Delta{(t)}/\Delta{(\nu^{-2}})$,
where $t(\nu)$ is the arrival time at frequency, $\nu$. A scaling
factor composed of physical and astronomical constants is needed
to convert $\mathcal{D}$ to DM. In the early days of pulsar astronomy
the relevant constants were defined to parts per million (ppm).  As
a result, a convention arose in which this conversion factor was
fixed. Over time, several  such conventions came about -- recipe
for confusion.  Meanwhile, over the past several years, the SI
system has been restructured  and the parsec is now exactly defined.
As a result, the present accuracy of the conversion factor is below
a part per billion --  many orders of magnitude better than the
best measurement errors of $\mathcal{D}$.  We are now in an awkward
situation wherein the primary ``observable", the DM, has incorrect
scaling factor(s).  To address these two concerns I propose that
astronomers report the primary measurement, $\mathcal{D}$ (with a
suggested normalization of $10^{15}\,$Hz), and not the DM.  Interested
users can convert $\mathcal{D}$ to DM without the need to know
secret handshakes of the pulsar timing communities.

\section{Motivation: confusion}
 \label{sec:Motivation}

UT 28 April 2020 was a memorable and fabulous day for the field of
magnetars and fast radio bursts (FRBs). On this day, CHIME
(400--800\,MHz) and STARE2 (1.2--1.5\,GHz) found  a burst towards
SGR\,1935+2154.  The CHIME project reported a fluence of a few
kJy\,ms\footnote{Revised three weeks later to $6\times 10^5$\,Jy\,ms}.
STARE2 reported a fluence of over $1.5\times 10^6\,{\rm Jy\,ms}$.
The STARE2 burst, if placed at the nearest FRB galaxy, could be
reasonably argued to be at the faint end of the FRB population and
thus a case was developed in which at least some FRBs  arise from
active magnetars.

The motivation for this note came from my experience in relating
the CHIME burst to the STARE2 burst.  A scaling constant, $K$ or
$a=K^{-1}$ (hereafter, ``$K|a$", using the Unix programming convention
where ``$|$" stands for OR), composed of fundamental physical and
astronomical constants, is need to extrapolate the arrival time at
one frequency given the arrival time at another frequency.  Being
uninitiated I computed this number using the latest constants. I
was unable to relate the two bursts. Eventually I learnt that, over
five decades go, the value of $K$ was ``fixed" with the slop taken
up by the dispersion measure. The hoary arcana of pulsar timing and
separately my frustrating experience provided me the impetus for
inquiry which, in due course, led to a number of interesting forays
into the revised SI system,  plasma physics and special relativity.
I thought the resulting study was of likely interest to astronomers
who are interested in pulsars and FRBs, hence this report.

This simple report is organized as follows. I start off by reviewing
the basic physics of propagation of radio waves in interstellar
plasma and summarize the literature in regard to $a|K$
(\S\ref{sec:DispersionRadioSignals}). Then, in
\S\ref{sec:FundamentalConstants}, I review the titanic changes that
have taken place in the definition of constants in astronomy (2012)
and in the SI system (2019).  In \S\ref{sec:ExactlyDMMeasuring} I
explore phenomena other than electrons which could contribute to
dispersive delay. The list includes ions, temperature of the plasma
(motion of electrons), ambient magnetic fields and relative motion
between the observer and interstellar plasma.  In view of this
situation there is little need for knowing {the {\it absolute} value
of the DM at the parts per million (ppm) level, let alone at the
parts per thousand (ppt) level. With the demonstration that the DM
is sensitive to a host of phenomena I suggest that we abandon the
DM as the primary observable that gets reported
(\S\ref{sec:ConclusionWayForward}). Instead, I urge radio astronomers
to report  the experimental measure, $\mathcal{D}$, which carries
no ideology or expectation with it. The DM can be straightforwardly
deduced from $\mathcal{D}$, to a precision only limited by that of
the physical constants (currently standing at under a part per
billion).

Below, unless stated otherwise, I will be using  the Gaussian
framework with associated CGS units -- the standard practice in
astronomy.

\section{Dispersion of Radio Signals}
	\label{sec:DispersionRadioSignals}

Electromagnetic pulses propagating through cold plasma obey the
dispersion relation
  \begin{equation}
	\omega^2 = \omega_{e}^2 + c^2k^2
		\label{eq:DispersionRelation}
  \end{equation}
where
 \begin{equation}
    \omega_{e}^2 = \frac{4\pi n_e e^2}{m_e}
	\label{eq:omega_e}
  \end{equation}
is the electron plasma angular frequency [B.\ Draine 2011, \S11.2].
For typical astrophysical conditions, $n_e$, the density of electrons
range from $10^{-4}\,{\rm cm^{-3}}$ to $10^4\,{\rm cm^{-3}}$ and
the electron plasma frequency $\nu_e=\omega_e/(2\pi)\approx 9(n_e/{\rm
cm^{-3}})^{1/2}\,$kHz, well below any conceivable observing frequency.
The  group velocity, $\nu_g$, is
	\begin{equation}
		v_g(\nu)\equiv\frac{\partial\omega}{\partial k} =
		c\Big(1-\frac{\omega_e^2}{\omega^2}\Big)^{1/2}
	\end{equation}
which leads to a frequency-dependent arrival time,
 \begin{eqnarray}
  t (\nu) &=& \int_0^L \frac{1}{v_g(\nu)}dl \label{eq:tnu_formal}\approx
   \int_0^L \frac{dl}{c}\Big(1 +
     \frac{1}{2}\frac{\omega_e^2}{\omega^2}\Big),\cr
      \label{eq:TaylorExpansion}
     \tau(\nu) &=& t(\nu) - t(\infty) = \frac{e^2\times{\rm
     parsec}}{2\pi m_e c}\frac{1}{\nu^2} {\rm DM}.
  \label{eq:tnu}
 \end{eqnarray}
Here, $L$ is the distance to the source, $t(\infty)=L/c$  and ${\rm
DM}=\int_0^L n_e dl$ with $n_e$ in cm$^{-3}$ and $L$ in pc. The
small value of $\epsilon=(\omega_e/\omega)^2\approx 10^{-10}\nu_9^{-2}n_e$
justifies the Taylor approximation made above; here, $\nu=10^9\nu_9\,$Hz.
There is a long history of astronomers searching for $\epsilon^2$
term [e.g., Goldstein \&\ James 1969; Tuntsov 2014], but with no
success.

From Equation~\ref{eq:tnu} we see that the DM is a product of $\nu^2$
and the dispersive delay to that frequency:
 \begin{equation}
	{\rm DM} = K \nu_9^2\tau(\nu)
	  \label{eq:DM_K}
 \end{equation}
where
 \begin{equation}
	a=K^{-1} \equiv\frac{e^2}{2\pi m_ec}\times {\rm parsec}.
  \label{eq:a}
 \end{equation}

The tradition of fixing the value of $K$ can be traced to Manchester
\&\ Taylor (1972): ``Because of the uncertainty in propagation
constants, most accurate published dispersions are quoted as the
dispersion constant, $DC = \Delta t/\Delta(1/f^2)$, which is a
directly measured quantity.  However, for consistency we have
converted these values to the more commonly quoted dispersion
measures, $DM$, using the relation $DM({\rm cm^{-3}\,pc}) =
2.41000\times 10^{-16}\,DC\,({\rm Hz}).$"

This suggestion took traction within some precision pulsar timing 
groups.  However, in the {\it Handbook of Pulsar
Astronomy}, a textbook  used by by students and researchers entering
the field of pulsars, the authors quote $a=4.148\,808(3)\ {\rm
GHz^2\,cm^{3}\,pc^{-1}\,ms}$ [D.\ R.\ Lorimer \&\ M.\ Kramer (2004)].
Some pulsar programs even use $a\equiv 4.15\ {\rm
GHz^2\,cm^{3}\,pc^{-1}\,ms}$.

Within a given school of pulsar timing the specific choice of $K|a$
does not matter since $K|a$ is fully covariant with the DM. However,
as explained in the previous section, difficulty will arise if one
uses the published DM without knowing the associated $K|a$.

\section{Fundamental Constants}
 \label{sec:FundamentalConstants}

More than five decades have passed since precision pulsar timing
began in earnest.  The situation in regard to physical constants
has substantially changed over this period. To start with, the IAU
in 2012, via Resolution B2,
fixed\footnote{https://www.iau.org/public/themes/measuring/} the
the value of AU. The same resolution defined the parsec to be the
small-angle-approximation\footnote{eschewing the proper trigonometric
formula!} distance to a star which subtends a parallax of an arcsecond
across an AU-wide baseline. Next,  on 2019 May 20 (the ``World
Metrology Day"), the Syst\`eme Internationale d$^\prime$Unit\'es
(SI)\footnote{\url{https://www.nist.gov/si-redefinition}} announced
permanent values for four fundamental constants\footnote{With this
announcement the seven defining constants of the  SI system are:
the hyperfine line frequency of Cesium-133 ($\Delta\nu_{\rm Cs}$),
$c$, $h$, $k_B$, $e$, Avogadro's number ($N_A$) and luminous efficacy
($K_{\rm cd}$). See \url{https://physics.nist.gov/cuu/Constants/index.html}
for values.}, which amongst other things, replaced the kilogram and
the ampere; see D.\ Newell [2014] for an overview. These two
developments provide additional impetus to investigate matters
related to precision pulsar timing.

Most astronomers routinely work within the Gaussian framework of
equations (and related constants). However, with the latest revision
to the SI system, some of the older SI to CGS conversions are no
longer valid. Next, in the revised SI system, the charge of the
electron, $e$, is exactly defined but in the Gaussian system, $e$
carries the uncertainty of the fine structure constant.  I  found
that converting from SI units to CGS was prone to errors and so I
will temporarily switch to the SI framework.  The electron plasma
frequency is then $\omega^2 = n_e e^2/(\epsilon_0 m_e)$ and so
 \begin{equation}
	K^{-1} = \frac{1}{8\pi^2}\frac{e^2}{\epsilon_0 m_e c}\times
	{\rm parsec}.
  \label{eq:KSI}
 \end{equation}
The SI constants which constitute $K$ are
 \begin{eqnarray*}
  c &\equiv& 299,792,458\,{\rm m\,s^{-1}},\\
   e &\equiv&  1.602\ 176\ 634\times 10^{-19}\,{\rm Coulomb},\\
    \epsilon_0 &=& 8.854\ 187\ 8128(13) \times10^{-12}\,{\rm
    F\,m^{-1}},\\
	 m_e &=& 9.109\ 383\ 7015(28)\times 10^{-31}\,{\rm kg},\\
	{\rm AU} &\equiv& 149,597,870,700\,{\rm m},\cr\cr
      {\rm parsec} &\equiv& \frac{180\times 3600}{\pi} {\rm AU}.
      \label{eq:constants}
 \end{eqnarray*}
Substituting these constants  into Equation~\ref{eq:KSI} and using
sensible normalizations (frequency in GHz, $n_e$ in cm$^{-3}$,
distance in parsec) I find
 \begin{eqnarray}
  a &=& 4.148\,806\,4239(11)\
			 {\rm GHz}^{2}\,{\rm cm^{3}\,pc}^{-1}\,{\rm
			 ms},\\
    K &=& 241.033\,1786(66)\
			{\rm GHz}^{-2}\,{\rm cm^{-3}\,pc}\,{\rm
			s}^{-1}.  \label{eq:aK}
	\label{eq:aK}
 \end{eqnarray}
Note that 8-byte floating point arithmetic should be used to preserve
the high precision of the constants.  The fractional uncertainty
in $K|a$ of $2.8\times 10^{-10}$ is a result of the experimental
fractional uncertainties of $m_e$ ($3\times 10^{-10}$)
and $\epsilon_0$ ($1.5\times 10^{-10}$).  This uncertainty is so
small, in relation to measurement errors of arrival times, that we
should no longer be setting $K|a$ to nice round numbers.

\section{What exactly is the DM  measuring?}
 \label{sec:ExactlyDMMeasuring}

Astronomers are interested in DM because it appears to be a highly
desirable physical quantity -- the column of electrons from the
source to the observer. However, a closer investigation shows that
other phenomena also contribute to the DM.  Some of these are
intrinsic (ions, temperature, magnetic field) and others are extrinsic
(relative motion). J.\ A.\ Phillips \&\ A.\ Wolszczan [1992] discuss
corrections arising from finite temperature of the interstellar
plasma and the effect of ambient magnetic fields on the inferred
DM. Below, I comment on this paper at appropriate points.

\subsection{Ions}
	\label{sec:Ions}

An external electromagnetic field induce motion of not only the electrons
but also the ions. Including both excitations leads to the following
dispersion relation:
	\begin{equation}
		\omega^2 = \omega_{e}^2 +\omega_{i}^2 + c^2k^2
	  \label{eq:ExactPlasmaFrequency}
	\end{equation}
where
	\begin{equation}
		\omega_{i}^2 = \frac{4\pi e^2}{m_p}\sum_{Z=1}\frac{n_Z
		q_Z^2}{A}
	\end{equation}
(see \S\ref{sec:DispersionWithIons}).  The sum is over ions of
atomic weight ($A$), atomic number ($Z$) and number density $n_Z$.
We assume that each atomic species is represented by its dominant
isotope and has only one dominant ionization state (charge $q_Z$).
Equation~\ref{eq:ExactPlasmaFrequency} can be restated as
	\begin{equation}
		\omega^2 =  \omega_p^2 + c^2k^2
	\end{equation}
where the ``plasma frequency" is
 \begin{eqnarray}
   \omega_p^2 &=& \frac{4\pi n_e e^2}{m_e}
    \Bigg[1+  \frac{m_e}{m_p}\sum_{Z=1}
    \frac{n_Z}{n_e}\frac{q_Z^2}{A}\Bigg]\cr
      &=& \omega_e^2\Bigg[1+  \frac{m_e}{m_p}\sum_{Z=1}
      \frac{n_Z}{n_e}\frac{q_Z^2}{A}\Bigg].
	\label{eq:omega_p}
 \end{eqnarray}

In effect,  Equation~\ref{eq:omega_p} informs us that the  the
square of the plasma frequency is equal to the charged particle
number density but with  each  particle weighted by  $Z^2/m$ where
$Z$ and $m$ is the charge and the mass of the particle, respectively.
Electrons dominate the sum because of their low mass.  The contribution
from protons is diminished by $m_e/m_p$ which amounts to 5 parts
in $10^4$ or 500 part per million (ppm).  In Table~\ref{tab:X} I
list the ``Cosmic" or Solar abundance of significant elements.

In the Galactic Warm Ionized Medium (WIM) Helium is not ionized and
so the dominant additional contribution is from protons.  Moving
on, the emerging view is that, at low redshift,  a significant fraction
of the baryons are distributed in an extended halo around galaxies
(the ``circumgalactic medium" or CGM; J.\ Tumlinson, M.\ Peeples
\& J.\ Werks  2017) as opposed to the traditional ``intergalactic
medium" (IGM; M.\  McQuinn 2016).  The IGM is heated by light from
active galactic nuclei whereas the CGM is heated by shocks generated
by infall. For the low redshift Universe ($z\lesssim 3$) we can
assume that Helium is fully ionized in the IGM in which case the
additional contribution is ${\rm X}({\rm He})m_e/m_p$, about 55 pm.
All metals, even if full ionized, will contribute, relative to
electrons, no more than a few ppm.

\begin{table}
 \begin{tabular}{rrrr}
  \hline
Z &   A &  Atom &  X \\
 \hline
  2   & 4   & He  &  $9.6\times 10^{-2}$   \\
   8   & 16  & O  & $5.4\times 10^{-4}$    \\
6  & 12   & C   &  $3.0\times 10^{-4}$     \\
 10  & 20  & Ne &  $9.3\times 10^{-5}$     \\
  7    &14   & N  & $7.4\times 10^{-5}$    \\ 
12  & 24  & Mg  &  $4.4\times 10^{-5}$     \\
 14 & 28   & Si   &  $3.6\times 10^{-5}$   \\
26  & 56  & Fe  &  $3.5\times 10^{-5}$     \\
  16  & 32  & S    &  $1.5\times 10^{-5}$  \\ 
 \hline
\end{tabular}
 \caption{\scriptsize The abundance of elements with atomic number
 $A$ and atomic charge $Z$ by number, relative to Hydrogen.  From
 Draine 2011, \S1.2.  }
  \label{tab:X}
\end{table}

\subsection{Warm Plasma}
	\label{sec:WarmPlasma}

The physics of warm plasma, $\epsilon_T=k_BT/(m_ec^2)\ll 1$, is not
only complex but is rife with  tedious algebra.  A simple and
qualitative understanding can be obtained by noting that the thermal
motion of electrons, rms velocity $v_e$, results in an increased
mass of the electrons, $\gamma_e m_e$; here,
$\gamma_e=(1-\beta_e^2)^{-1/2}$ with $\beta_e=v_e/c$.  As can be
seen from Equation~\ref{eq:omega_e} a heavier electron leads to a
smaller dispersive delay. The mean of the inverse mass of the
electron is decreased by $\langle\gamma\rangle^{-1}\approx
1-(1/2)\langle\beta_e^2\rangle$ where the averaging is done over
the Maxwellian distribution. Since
$\langle\beta_e^2\rangle=3kT_e/(m_ec^2)$, the fractional contribution
amounts to $O(-\epsilon_T)$.

Buneman (1980) derives dispersion relation for warm plasma,
$\epsilon_T\ll 1$.   Further simplifying  this relation for low
density plasma, $\epsilon \ll 1$, I find
 \begin{equation}
   \omega^2 = k^2 c^2 + \Big[1- {\textstyle
   \frac{3}{2}}\epsilon_T\Big]\omega_e^2
  \label{eq:WarmDispersion}
 \end{equation} (see \S\ref{sec:WarmPlasmaDispersion}).
Equation~\ref{eq:WarmDispersion} amazingly agrees with the simpler
argument presented above.  This correction amounts to {\it decreasing}
the contribution by electrons by  $f=-\frac{3}{2}\epsilon_T$.  The
result presented here differs from Phillips \& Wolszczan [1992]
who, with no justification, state  $f=+\epsilon_T$.

Since $f\approx -253T_6$\,ppm the breakeven temperature is $2\times
10^6\,$K.  At this temperature, the increased mass of the moving
electrons compensates for the contribution to the dispersive delay
by protons. For even hotter gas, say cluster gas with $T\approx
5\times 10^7\,$K, the hot electrons will decrease the dispersive
delay by up to $-1\%$.

\subsection{Magnetic Fields}
	\label{sec:MagneticFields}

The dispersion relation for electromagnetic waves through magnetized
plasma is
	\begin{equation}
		\omega^2 = k^2c^2 + \frac{\omega_e^2}{1\pm
		(\omega_B/\omega)}
	\end{equation}
where $\omega_B=eB/(m_e c)$ is the electron gyro-frequency and the
$\pm$ applies to the two senses of circular polarization [Draine
2011, \S11.3].  For ISM parameters,  $\eta=\omega_B/\omega\ll 1$,
and so the above relation can be  simplified to
 \begin{equation}
   \omega^2 = k^2c^2 + \omega_e^2\Big(1\mp \frac{\omega_B}{\omega}\Big)
 \end{equation}
leading to a dispersive delay that is both frequency and polarization
dependent:
 \begin{equation}
	\tau(\nu) = a\frac{\rm DM}{\nu^2} \Bigg[1 \pm
	2\Big(\frac{\nu_B}{\nu}\Big)\Bigg].
 \end{equation}
The fractional change in the dispersive delay is $4\eta= 1.2\times
10^{-8}B_\mu\nu_9^{-1}$ with  $B_{\mu}$ being the magnetic field
strength along the line-of-sight and in $\mu$G.  Phillips \&\
Wolszczan [1992] reach a similar conclusion.

This effect is only important  if the intervening cloud is highly
magnetized (perhaps a cloud local to FRB) and that too at low
observing frequency.  For instance, if the local magnetization is
1\,mG then the effect is $10^{-4}(\nu/100\,{\rm MHz})$. This effect
would be diluted if most of the contribution to the DM came from
gas in ISM or IGM.

\subsection{Relative Motion}
	\label{sec:RelativeMotion}

The motion of the pulsar with respect to the intervening medium or
the observer does not affect the inferred DM. However, the relative
motion of the observer with respect to the intervening medium will
lead the observer to infer a different value for  the DM.

Consider an event which puts out a broadband pulse, radio through
X-ray.  We start by considering the simple case of an intervening
plasma cloud that is stationary with respect to an observer located
at the solar system barycenter (SSB).  Owing to dispersive delay
within the cloud, the radio pulse, frequency $\nu_0$, arrives
$\tau_i$ after the X-ray pulse (which we assume traveled at the
speed of light).  Following Equation~\ref{eq:DM_K} the observer
infers
  \begin{equation}
		{\rm DM}_i= K\tau_i\nu_0^2.
	\label{eq:DM_i}
 \end{equation}

Next, consider the case of an observer stationed on Earth. The
orbital velocity of Earth around the SSB can range up to $\pm
30\,{\rm km\,s^{-1}}$.  The frequency of the radio pulse as perceived
by the Earth-bound observer is given by the relativistic Doppler
formula:
	\begin{equation}
		\nu = \frac{1}{\gamma}\Big(\frac{1}{1+\beta_r}\Big)\nu_0
		\label{eq:DopplerExact}
	\end{equation}
where $v_r=c\beta_{\rm r}$ is the radial velocity\footnote{I follow
the astronomical convention in which receding radial velocities are
positive.}, $\beta=v_{\rm orb}/c$ and $\gamma=(1-\beta^2)^{-1/2}$.

Let us assume that we have arranged for a light pulse to be emitted
when the radio pulse enters and exits the cloud. All quantities,
unless stated otherwise, are in the frame of the observer. Let the
distance from the observer to the cloud at the time of entry be
$L$.  The first light pulse arrives at time $t_1=L/c$. From time
dilation we know that the second pulse will be emitted $\Delta
t=\gamma\tau_i$ (observer frame) later. Thus, the observer receives
the second light pulse at time $t_2=\Delta t+(L+v_r\Delta t)/c$.
As a result, the dispersive delay measured by  the observer, $t_2-t_1$,
is $\tau=(1+\beta_r)\gamma\tau_i$.  As before (Equation~\ref{eq:DM_K})
the inferred DM is given by the product of the dispersive delay and
square of the frequency, both measured in the same frame, or
 \begin{equation}
   {\rm DM}= K\tau\nu^2= \frac{1}{\gamma}\frac{1}{(1+\beta_r)}{\rm
   DM}_i.
	\label{eq:DM_DMi}
 \end{equation}
Thus, the observer value of DM is the intrinsic value times the
Doppler factor.  It appears that the  
DM behaves like a spectral line.  In cosmology,
without large-scale structure, all motions are radial ($\beta_r=\beta$)
and so the observed DM is $(1+z)$ smaller than the intrinsic value
-- a well known result [K.\ Ioka 2003, S.\ Inoue 2004].

Since $\beta_r$ is small we see that the fractional error in the
inferred DM is $\pm \beta_{\rm r}$.  For routine observations such
as searches the variation of DM with respect to SSB is not taken
into account. However, precision timing programs such as \texttt{tempo}
and \texttt{tempo2} are quite aware of Equation~\ref{eq:DM_DMi}
(M.\ Bailes, pers.\ comm.).

Incidentally, FRB observers at two different Earth-based facilities
observing the same burst will perceive slightly different frequencies,
owing to differing (Earth) rotational velocities. The resulting
fractional difference in the inferred DM is smaller than  $v_{\rm
rot}/c\approx 3\times 10^{-6}$ or 3 parts per million. Parenthetically
we note that the ionospheric vertical electron column density varies
between 5 to 500\,TEC\footnote{The TEC or ``total electron column"
is a unit used by aeronomers and is equal to a column density of
$10^{12}\,{\rm cm^{-2}}$; see \url{http://solar-center.stanford.edu/SID}.}
depending on diurnal and sunspot phase. Likely, for almost all FRBs,
the signal-to-noise ratio will not be high enough to make a material
difference to this discussion.

Equation~\ref{eq:DM_DMi}, thanks to the principle of relativity,
equally applies for the case of the cloud moving with respect to
the SSB.  Notice that the fractional correction to DM has the sign
of the radial velocity.  So there will be a reduction of this effect,
should the ray go through several clouds with opposing radial
velocities.  However, the second order corrections still remain at
the level of  $O(\beta^2)$.

Given cosmographical parameters and assuming that most of the baryons
are in the IGM  a formal formula can be written down as a a function
of $(1+z)$  [K.\ Ioka 2003, S.\ Inoue 2004]. Convenient fitting
formula are readily available [e.g.\ Z. Zheng et al.\ 2014].  However,
for any given FRB, owing to large- and small-scale structure in
(baryonic and dark) matter, significant deviations in DM, with
respect to such formulae, are expected.  The deviations will depend
on the number of times the line-of-sight crosses clusters of galaxies
and intersects CGM halos of galaxies; see M.\ McQuinn [2017].

From Equation~\ref{eq:DM_DMi} we see that peculiar velocities
(velocities deviating from pure Hubble radial flow) will result in
additional but minor perturbations.  Examples of peculiar velocities
include the rotation curve of the Milky Way (250\,km\,s$^{-1}$),
the infall of our  Galaxy towards M31 (100\,km\,s$^{-1}$) and the
peculiar velocity due to structure on the local
Baryon Acoustic Oscillation (BAO) scale (about 400\,km\,s$^{-1}$).
The latter is most elegantly measured by observations of the Cosmic
Background Radiation (CMB) ``dipole" [e.g.\ Planck Collaboration
2014].  For gas at higher redshift all such kinematic effects are
suppressed by $(1+z)$. I wonder,  whether in  a decade from now,
when FRBs are routinely localized to arcsecond accuracy every hour
(and redshifts of host galaxies already determined from
massively-multiplexed spectroscopic surveys), we will be able to
sense the CMB dipole via this effect.

\subsection{Deviations from $\nu^{-2}$}

Above we have assumed that the pulsed signal arrives strictly as
$\nu^{-2}$.  However, there are indications of increased timing
noise at low frequencies and one of the possibilities is differences
in path taken by rays at low and high frequencies [R.\ M.\ Shannon
\&\ J.\ M.\ Cordes 2017].  Low frequency timing has its own additional
opportunities and complications. This topic is beyond the scope of
the paper.

\section{Conclusion \&\ Way Forward}
	\label{sec:ConclusionWayForward}

\subsection{Accurate Dispersion Measure: (F)utility}
 \label{sec:Futility}

PSR~1909-3744  (DM=10.4\,cm$^{-3}$\,pc) probably has the most precise
DM measurement, about $10^{-5}\,{\rm cm^{-3}\,pc}$ per epoch [M.\
L.\ Jones et al. 2017].  Most pulsars, when carefully monitored,
show changes in the DM at the level of $10^{-4}\,{\rm
cm^{-3}\,pc\,yr^{-1}}$ [e.g.\ V.\ M.\ Kaspi, J.\  H.\ Taylor \&\
M.\ F.\ Ryba 1994;  Lam et al.\ 2016].  Such annual changes at the
level of 1 to 100 ppm primarily arise from secular evolution of the
pulsar-observer line of sight and/or gradual changes in the ISM
itself.  Furthermore, as discussed earlier in the paper, the DM,
in addition to being the measure of the electron column density,
is also sensitive to protons (500 ppm), motions of intervening
clouds ($10$ to $10^3$ ppm) and the temperature (up to $-1$\%) of
the intervening gas.

Thus, for most purposes, an accurate value of DM is not all that
useful. On the other hand, astronomers undertaking precision pulsar
timing have to account for changes in DM and for this purpose the
measured quantity
 \begin{equation}
	\mathcal{D}(\nu_1,\nu_2)\equiv\frac
	{t(\nu_1)-t(\nu_2)}{\nu_1^{-2}-\nu_2^{-2}}
  \label{eq:mathcalD}
 \end{equation}
is a sensible and apt quantity. $\mathcal{D}$ carries the unit of
Hz.  A convenient normalization is $\mathcal{D}$ is $10^{15}\,{\rm
Hz}$ since it corresponds to about $0.4\,{\rm cm^{-3}\,pc}$. Thus,
I suggest that $\mathcal{D}_{15}\equiv\mathcal{D}\times 10^{-15}\,$Hz
be recorded and reported.

Separately,  colloquially it is often stated that ``the dispersion
measure is the column density of electrons". One also hears of
frequent press announcements proclaiming that ``FRBs allow astronomers
to count every electron along the line of sight".  However, as
extensively discussed in \S\ref{sec:Ions} and \S\ref{sec:WarmPlasma},
the DM is sensitive to ions as well and has temperature dependence.
Furthermore, unlike true column densities, the DM is not a Lorentz
invariant (\S\ref{sec:RelativeMotion}).  We conclude that the DM
is a good proxy for the electron column density for routine
astronomical purposes, say to one ppt, but highly accurate measurements
of the DM, say to one ppm, do not carry proportionally valuable
information.

\subsection{A Way Forward}

In view of the arguments presented in the previous section I suggest
that we abandon DM as one of the key precision parameters of FRBs
and pulsars.  Instead, I urge my colleagues to report
$\mathcal{D}(\nu_1,\nu_2)$ for both FRBs and pulsars.  Since
$\mathcal{D}$ is not a Lorentz invariant it is essential to report
the topocentric and barycentric values.  I note that Table~1 of R.\
N.\ Manchester [1971]  is a fine example for reporting $\mathcal{D}$
(followed by the DM).

Given $\mathcal{D}$, astronomers can  readily obtain the DM from
	\begin{equation}
		{\rm DM} =  K\mathcal{D}
	\end{equation}
with the full assurance that $K$ (Equation~\ref{eq:aK}) is known
to better than one part per billion.  For almost all purposes that
I can think of there is little need for {\it accuracy} of DM, say,
beyond even a part per thousand.  Neophytes can compute the dispersive
delay to any frequency via the equation $\tau(\nu)=a\mathcal{D}\nu^{-2}$.
Fastidious users can apply appropriate corrections,  both Special
relativistic and General relativistic, to $\mathcal{D}$.

The old name for $\mathcal{D}$ was ``dispersion constant" [see R.\
N.\ Manchester \&\ J.\ H.\ Taylor 1972].  However, $\mathcal{D}$
is not a time invariant for a given pulsar, being affected by the
secular evolution of the line-of-sight from the observer to the
pulsar and gradual changes in the ISM (\S\ref{sec:Futility}).
Obviously different pulsars have different values of $\mathcal{D}$.
In view of this I suggest the term ``Dispersion Slope" for
$\mathcal{D}$.

The proposal made here has the distinct advantage of preserving the
precision of the measured quantity, $\mathcal{D}$ (or $\mathcal{D}_{15}$),
in {\it published} literature. Perhaps, equally importantly, this
proposal will result in sparing tyros, attempting to link the burst
at different frequencies, from the need to learn secret handshakes
of pulsar timing clubs.

\smallskip

\noindent{\bf Acknowledgements.} I am grateful to Wenbin Lu and E.\
Sterl Phinney, III for  extensive discussions and considerable help.
I thank Matthew Bailes,  Ilaria Caiazzo, Joe Lazio, Michael Kramer,
Vikram Ravi, Marten van Kerkwijk and Harish Vedantham for discussions.
I am indebted to Ravi  for acting as  an internal referee.

\bigskip \noindent{\bf References}

\noindent O.\ Buneman, Astrophysical Journal, 235, 616
pp.\ (1980)

\noindent F.\ F.\ Chen, ``Introduction to Plasma Physics", Plenum
Publishing, (1974)

\noindent B.\ T.\ Draine, ``Physics of the Interstellar \&\
Intergalactic Medium", Princeton University Press, (2011, sixth
printing)

\noindent S.\ Inoue, Monthly Notices of the Royal Astronomical
Society, 348, 999 pp.\ (2004)

\noindent K.\ Ioka, Astrophysical Journal (Letters), 598, L79 pp.\
(2003)

\noindent V.\ M.\ Kaspi, J.\ H.\ Taylor \&\ M.\ F.\ Ryba,
Astrophysical Journal, 428, 173 pp. (1994)

\noindent M.\ L.\ Jones et al., Astrophysical Journal, 841, 21 pp.\
(2017)

\noindent M.\ T.\ Lam et al., Astrophysical Journal, 821, 1 pp.\
(2016)

\noindent R.\ N.\ Manchester, Astrophysical Journal (Letters), 167,
L101 pp.\ (1971)

\noindent R.\ N.\ Manchester \&\ J.\ H.\ Taylor, Astrophysical
Letters, 10, 67 pp.\ (1972)

\noindent M.\ McQuinn, Astrophysical Journal (Letters), 780, L33
pp.\ (2014)

\noindent M.\ McQuinn, Annual Review of Astronomy \&\ Astrophysics,
54, 313 pp.\ (2016)

\noindent D.\ B.\ Newell, Physics Today, 67, 35 pp.\ (2014)

\noindent Planck Collaboration, Astronomy \&\ Astrophysics, 571,
27 pp.\ (2014)

\noindent J.\ A.\ Phillips \&\ A.\ Wolszczan, Astrophysical Journal,
385, 273 pp.\ (1992)

\noindent R.\ M.\ Shannon \&\ J.\ M.\ Cordes, Monthly Notices of the
Royal Astronomical Society, 464, 2075 pp.\ (2017)

\noindent J.\ Tumlinson, M.\ S.\ Peeples \&\ J.\ Werk, Annual Review
of Astronomy \&\ Astrophysics, 55, 1 pp.\ (2017)

\noindent A.\ V.\ Tuntsov, Monthly Notices of the Royal Astronomical
Society, 441, L26 pp.\ (2014)

\noindent Z.\ Zheng et al., Astrophysical Journal, 797, 71 pp.\
(2014)


\appendix

\section{Dispersion relation including the contribution from ions}
\label{sec:DispersionWithIons}

F.\ F.\ Chen [1974; \S4.12] provides the starting point for this
section.  As an external electromagnetic field propagates through
the plasma it will excite small currents ${\bf j}_1$ which in turn
generate electromagnetic fields. We will assume that there is no
external magnetic field threading the plasma ($B_0=0$) and also
that the plasma is ``cold" (no gas pressure).

The relevant Maxwell's equations for the electromagnetic fields are
  \begin{eqnarray}
	\nabla\times{\bf  E_1} &=& -\frac{1}{c}\dot {\bf B}_1
		\label{eq:E_Maxwell},\\
	\nabla\times{\bf B_1} &=& \frac{1}{c}\dot{\bf E}_1 +
	\frac{4\pi}{c}{\bf j}_1.
		\label{eq:B_Maxwell}
  \end{eqnarray}
The curl of Equation~\ref{eq:E_Maxwell} is
  \begin{eqnarray*}
   \nabla\times(\nabla\times{\bf E}_1)&=&\nabla(\nabla\cdot{\bf
   E}_1)-\nabla^2{\bf E}_1 =-\frac{1}{c}\nabla\times\dot{\bf B}_1
  \end{eqnarray*}
whereas the time derivative of Equation~\ref{eq:B_Maxwell} is
  \begin{eqnarray*}
	\nabla\times \dot{\bf B}_1 &=& \frac{1}{c}\ddot{\bf E}_1+
	\frac{4\pi}{c}\dot{\bf j}_1.
   \end{eqnarray*}
Eliminating  $\nabla\times\dot{\bf B}_1$ between the last two
equations and letting ${\bf E}_1\propto \exp(i{\bf
k}\cdot{\bf x} -i\omega t)$ yields
  \begin{eqnarray*}
	-{\bf k}\cdot({\bf k}\cdot{\bf E}_1)+k^2{\bf E}_1 &=&
	\frac{4\pi i\omega}{c^2} {\bf j}_1
		+ \frac{\omega^2}{c^2}{\bf E}_1.
  \end{eqnarray*}
Electromagnetic waves are transverse waves and so ${\bf k}\cdot{\bf
E}_1=0$. Thus,
  \begin{equation}
	(\omega^2 -c^2k^2){\bf E}_1=-4\pi i\omega {\bf j}_1.
	\label{eq:omega2}
   \end{equation}
In the absence of plasma, the RHS is zero and we would then find
$\omega^2=k^2c^2$, as expected.

The electric current is due to electrons and ions
  \begin{equation}
	{\bf j}_1=-n_e e {\bf v}_{e1}+ n_i e q_i {\bf v}_{i1}
	 \label{eq:j1}
  \end{equation}
where $n_e$ and ${\bf v}_{e1}$ is the number density and velocity
of the electrons and $n_i$, ${\bf v}_{i1}$ and $q_ie$ is the number
density, velocity and charge of the ions.  We use the fluid mechanics
equation of force to compute the velocities:
	\begin{eqnarray}
		m_e\Big[\frac{{\partial \bf v}_{e1}}{\partial t}
		+({\bf v}_{e1}\cdot\nabla){\bf v}_{e1}\Big] &=&
		-e\Big[{\bf E}_1 + \frac{{\bf v}_{1e}}{c}\times
		{\bf B}_1\Big].
	\end{eqnarray}
In the small amplitude approximation we only retain terms which are
linear in perturbed quantities. The second term in the LHS, being
$O(v_{e1}^2)$, can be ignored.  We now turn to the RHS. To start
with note that there is no pressure term on the RHS, consistent with
the assumption of cold plasma. Next, as can be deduced from
Equation~\ref{eq:E_Maxwell}, the strength of $E_1$ is similar to
that of $B_1$. However, the force due to magnetic field is reduced
by $v_{e1}/c$. In our small amplitude approximation, the velocities
of the fluid are small relative to the speed of light, $v_{e1}/c\ll
1$. Thus, in this approximation the force on the electron
due to (propagating) magnetic fields can be neglected.  The result is
 \begin{eqnarray*}
	m_e\frac{\partial{\bf v}_{e1}}{\partial t} = -e {\bf
	E}_1,\qquad && m_i\frac{\partial{\bf v}_{i1}}{\partial t}
	= q_ie {\bf E}_1.
 \end{eqnarray*}
The electron and ion velocities are excited by the incident
electromagnetic field and thus they should also have the same
functional form or ${\bf v}_{e1}\propto \exp[i{\bf k}\cdot{\bf
x}-i\omega t]$ and so
  \begin{eqnarray*}
	{\bf v}_{e1} =  \frac{e{\bf E}_1}{im_e\omega}, \qquad &&
	{\bf v}_{q1} = -\frac{qe{\bf E}_1}{im_i\omega}.
		\label{eq:v}
  \end{eqnarray*}
Substituting these velocities into Equation~\ref{eq:j1} we find
that the total current is
  \begin{equation}
	{\bf j}_1 =-\frac{1}{i\omega}\bigg[\frac{n_ee^2}{m_e}+
	\frac{n_iq_i^2e^2}{m_i}\bigg]{\bf E}_1
  \end{equation}
which, when substituted into Equation~\ref{eq:omega2}, leads to the
dispersion law:
  \begin{equation}
	\omega^2 = k^2c^2 + \frac{4\pi n_e e^2}{m_e}+\frac{4\pi
	n_iq_i^2e^2}{m_i}.
  \end{equation}
As can be seen from Equation~\ref{eq:v} it is easy to extend the
dispersion relation to include contributions from several species
of ions.

\section{Dispersion Relation for Warm Plasma}
 \label{sec:WarmPlasmaDispersion}

O.\ Buneman [1980] provides the following  dispersion relation of
warm plasma which is accurate to  $O(\xi)$ where $\xi=\langle
v^2\rangle/c^2$ and $\langle v^2\rangle$ is thermal velocity
dispersion of the electrons in the plasma:
 \begin{equation}
	\frac{\omega^2}{\omega_e^2} =
	\frac{1}{1-n^2}\Big(1-\frac{\xi}{2}\Big)-\frac{\xi}{3}.
   \label{eq:Buneman}
 \end{equation}
Here,  $\omega$ is the angular frequency; $k$, the wavenumber;
$\omega_e$ the electron plasma angular frequency; $n=c/v_p$ is the
refractive index where the phase velocity is $v_p=\omega/k$.

Let, as in the main text, $\epsilon=\omega_e^2/\omega^2$.  For our
purpose, we seek to further simplify Equation~\ref{eq:Buneman} for
the case of $\epsilon\ll 1$. Noting that $n=kc/\omega$,
Equation~\ref{eq:Buneman} can be recast as
 \begin{equation*}
   \frac{\omega^2}{\omega_e^2}\Big(1-\frac{k^2c^2}{\omega^2}\Big)
    =
    \Big(1-\frac{\xi}{2}\Big)-\frac{\xi}{3}\Big(1-\frac{k^2c^2}{\omega^2}\Big)
 \end{equation*}
which can be re-arranged to yield
 \begin{eqnarray*}
	\omega^2-k^2c^2-\omega_e^2\Big(1-\frac{5}{6}\xi\Big) &=&
	\frac{1}{3}\xi\epsilon k^2c^2.
  \label{eq:formal_dispersion_law}
 \end{eqnarray*}
We see that Equation~\ref{eq:formal_dispersion_law} simplifies to
the usual plasma dispersion law when $\xi=0$.  Further rearranging
whilst dropping $O(\epsilon^2)$ results in
 \begin{eqnarray}
	k^2c^2 &=& \omega^2
	\Big[1-\epsilon(1-\textstyle{\frac{5}{6}}\xi)\Big]\Big/\Big[1+
	\textstyle{\frac{1}{3}}\xi\epsilon\Big]\cr
	&\approx&\omega^2\Big[1-\epsilon\Big(1-
	\textstyle{\frac{5}{6}}\xi\big)\Big]\Big[1-\textstyle{\frac{1}{3}}\xi\epsilon\Big]\cr
	&=& \omega^2\Big[\textstyle{1-\epsilon+\frac{1}{2}\epsilon\xi
	+ \frac{1}{3}\xi\epsilon^2 -\frac{5}{18}\xi^2\epsilon^2}\Big]\cr
	&\approx&\omega^2
	-\omega_e^2\Big[\textstyle{1-\frac{1}{2}\xi}\Big].
  \label{eq:k2c2}
 \end{eqnarray}

\end{document}